\documentclass[prl,twocolumn,showpacs,superscriptaddress,floatfix]{revtex4}

\usepackage{graphicx}
\usepackage{color}
\usepackage{mathrsfs}

\begin{document}

\title{Spin density wave dislocation in chromium probed by coherent x-ray diffraction}

\author{V.L.R. Jacques}
\affiliation{Laboratoire de Physique des Solides (CNRS-UMR 8502), B{\^a}t. 510, Universit{\'e} Paris-sud, 91405 Orsay cedex, France}\affiliation{Synchrotron SOLEIL - L'Orme des Merisiers Saint-Aubin - BP 48 91192 GIF-sur-YVETTE, France}
\author{D. Le Bolloc'h}
\affiliation{Laboratoire de Physique des Solides (CNRS-UMR 8502), B{\^a}t. 510, Universit{\'e} Paris-sud, 91405 Orsay cedex, France}
\author{S. Ravy}
\affiliation{Synchrotron SOLEIL - L'Orme des Merisiers Saint-Aubin - BP 48 91192 GIF-sur-YVETTE, France}
\author{C. Giles}
\affiliation{Instituto de Física "Gleb Wataghin", UNICAMP, Campinas-SP, C.P. 6165, 13083-970, Brazil}
\author{F. Livet}
\affiliation{LTPCM (CNRS-UMR 5614), ENSEEG, BP 75, 38402 Saint Martin d'Hères cedex, France}
\author{S.B. Wilkins}
\affiliation{Brookhaven National Lab, Upton NY 11973, USA}

\begin{abstract}
We report on the study of a magnetic dislocation in pure chromium.
 Coherent x-ray diffraction profiles obtained on the incommensurate Spin Density Wave (SDW) reflection are consistent with the presence of a dislocation of the magnetic order, embedded at a few micrometers from the surface of the sample. 
Beyond the specific case of magnetic dislocations in chromium, this work may open up a new method for the study of magnetic defects embedded in the bulk.  
\end{abstract}
\maketitle
\section{Introduction}
\label{intro}
Topological defects such as dislocations play an important role in the physics of condensed matter \cite{Friedel}.
They are easily observed by transmission electron diffraction or x-ray topography.
Purely electronic orderings such as Charge Density Waves (CDW) may also display their own defects. While these electronic dislocations were predicted 30 years ago \cite{lee,Gor'kov,maki,maki2,FF}, only a few experiments have observed
 such defects. Discommensurations 
have been reported by using dark field electron microscopy in 2D systems like 2h-TaSe$_2$ \cite{chen} and CDW dislocations have been seen by STM~\cite{cbrun} on a surface. CDW dislocations embedded a few micrometers from the surface were recently observed by using coherent x-ray diffraction~\cite{RefJ1}.
 
In magnetic systems, several techniques exist to image domains. For example, optical methods based on the Kerr effect \cite{Metaxas},
soft x-ray techniques based on imaging the photoelectrons, atomic force microscopy and spin polarized STM.
In chromium, for example, spin polarized STM measurements have shown that
screw dislocations are visible at the Cr(001) surface, leading to the formation of magnetic domains \cite{kleiber}. 
All these techniques are limited to surface studies or to the first tenth atomic layers.

Neutrons and magnetic x-ray diffraction remain the two prime techniques to probe $bulk$ magnetic long range order.
However, standard x-ray or neutron scattering experiments are based on a space averaging from which it is difficult to 
obtain precise information about dislocations. This can be overcome by illuminating the sample with a coherent beam of x-rays. Recent experiments using coherent soft x-rays (770-780 eV) showed that it is possible to track local reversal processes in magnetic nanostructures~\cite{beutier}. 
However, due to the high absorption of soft x-rays, only the surface of the sample was probed. 
In this paper we show that magnetic defects, embedded several micrometers in the bulk, 
can be probed by non-resonant coherent x-ray diffraction at a higher energy ($\sim 6$ keV).

\section{Spin versus charge density waves}
\subsection{Charge and spin density wave in quasi 1D systems}
\label{sec:1}
While charge and spin density waves present some similarities, their physical properties are very different. The former is a spatial modulation of the charge density with no resultant magnetization, whereas the latter is a spatial modulation of the spin density, with a constant charge density. However, 
 surface nesting plays a fundamental role both for the formation of a CDW and a SDW. Periodic lattice distortions are responsible for the formation of CDWs, whereas
 SDWs arise as a consequence of electron-electron interactions. But in both cases, the electronic energy 
is lowered by a gap opening in the dispersion curve at the Fermi wave vector $\pm k_F$. This theory has been developed 
for quasi-one-dimensional systems, but more generally, can be applied to any system displaying Fermi surface nesting.

In the case of CDWs, an electron-phonon interaction is responsible for the so-called Peierls 
lattice distortion, which induces a modulation of the charge density $\rho(x)$ in the crystal:
$$\rho(x)=n_0+\frac{2\Delta}{\Lambda}\cos(2k_Fx+\phi),$$
where $n_0$ is the average electron density and $\Lambda$ a constant that expresses the coupling strength between the lattice and the electrons. 
The interaction couples states at the Fermi surface separated by the nesting wave vector 
q=$\pm2k_F$ as expressed in the electron-hole polarisability \cite{Gruner,Pouget}:
$$\chi(q)=\sum_k\frac{<n_{k+q}>-<n_k>}{\varepsilon_k-\varepsilon_{k+q}},$$
where $n_k$ is the occupation number of the state $k$. The cost in terms of energy of the distortion
 is always less significant than the gain of electronic energy below a certain temperature
 in a one-dimensional system. A gap opens at the Fermi level, and the crystal undergoes a metal-insulator transition. 
The order parameter $\Delta$ is directly given by the amplitude of the lattice distortion and the Peierls temperature
 $T_P$ is given by the universal BCS ratio:
$$\frac{\Delta(0)}{k_BT_P}=1.76$$
where $\Delta(0)$ is the value of the gap at $T=0K$.

In the case of SDWs, the electron-electron interaction drives the instability. 
An electron and a hole separated by the nesting wave vector are coupled, and it can be shown 
that a SDW develops below a certain temperature in a one-dimensional system.
 Although no electron-phonon interaction occurs, contrary to the case of CDW, the polarisability takes the same expression as the one found in a CDW system. Moreover, the transition temperature is given by the same BCS ratio, and the same explanation 
of the instability with the nesting properties of the Fermi surface can be developed. 
The relationship between CDWs and SDWs becomes obvious when one considers the superposition of two in-phase or out-of-phase spin-polarized CDWs with opposite spins:
\begin{eqnarray*}
\rho_\uparrow(z)&=&\rho_0\left[1+\frac{\Delta}{V_Fk_F\Lambda}\cos{(2k_Fz)}\right]\\
\rho_\downarrow(z)&=&\rho_0\left[1+\frac{\Delta}{V_Fk_F\Lambda}\cos{(2k_Fz+\phi')}\right]
\end{eqnarray*}
where $V_F$ is the Fermi velocity and $\Lambda$ the constant containing the electron-electron coupling.
Taking a phase $\phi'=0$ leads to the formation of a CDW and $\phi'=\pi$ describes a SDW. Intermediate $\phi'$ values can be found in some compounds, leading to a mixed CDW and SDW state \cite{PougetRavy}.\\

\subsection{SDW and CDW in chromium}
\label{sec:12}

\begin{figure}[!ht]
$$\resizebox{1\columnwidth}{!}{%
\includegraphics[width=\columnwidth]{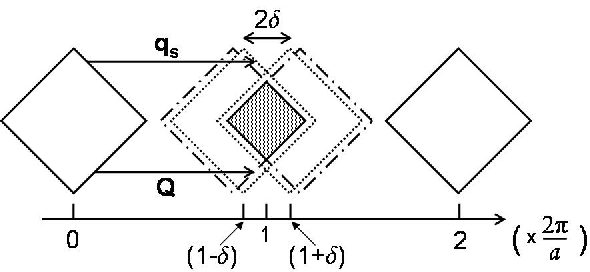}

}$$
\caption{Schematic Fermi surface section and nesting along one reciprocal direction. 
The grey region is an electron pocket, the white regions closed by solid lines are hole pockets. 
The SDW instability at $\frac{2\pi}{a}(1\pm\delta)$ and the CDW instability at $\frac{2\pi}{a}(\pm2\delta)$ are displayed. 
Dashed dot lines and dotted lines are the hole pockets translated by $\vec{Q}$ or $\vec{q_s}$ respectively (see text).}
\label{surf_fermi}       
\end{figure}

The case of chromium is quite unusual. Despite the absence of a one-dimensional structure, 
the Fermi surface exhibits a nesting property 
along the <100> directions (the calculated Fermi surface of chromium can be found in \cite{fry}). 
The Fermi surface consists of an electron pocket and two slightly larger hole octahedrons, 
and is shown schematically in figure~\ref{surf_fermi}. The imperfect nesting between electron
 and hole pockets (attractive electron-hole interaction) gives rise to an incommensurate SDW in chromium. 
 Note that theoretical work~\cite{fishman} shows that the SDW wave vector is not
 obtained exactly at the geometrical nesting vector $\vec{Q}$, because the nesting is actually maximized 
for a slightly different wave vector $\vec{q_s}$ (see figure \ref{surf_fermi}). This effect leads to the existence
 of particular magnetic excitations in chromium, termed wavons~\cite{theory}, and will be mentioned later.

\begin{figure}[!ht]
$$\resizebox{1\columnwidth}{!}{%
\includegraphics[width=\columnwidth]{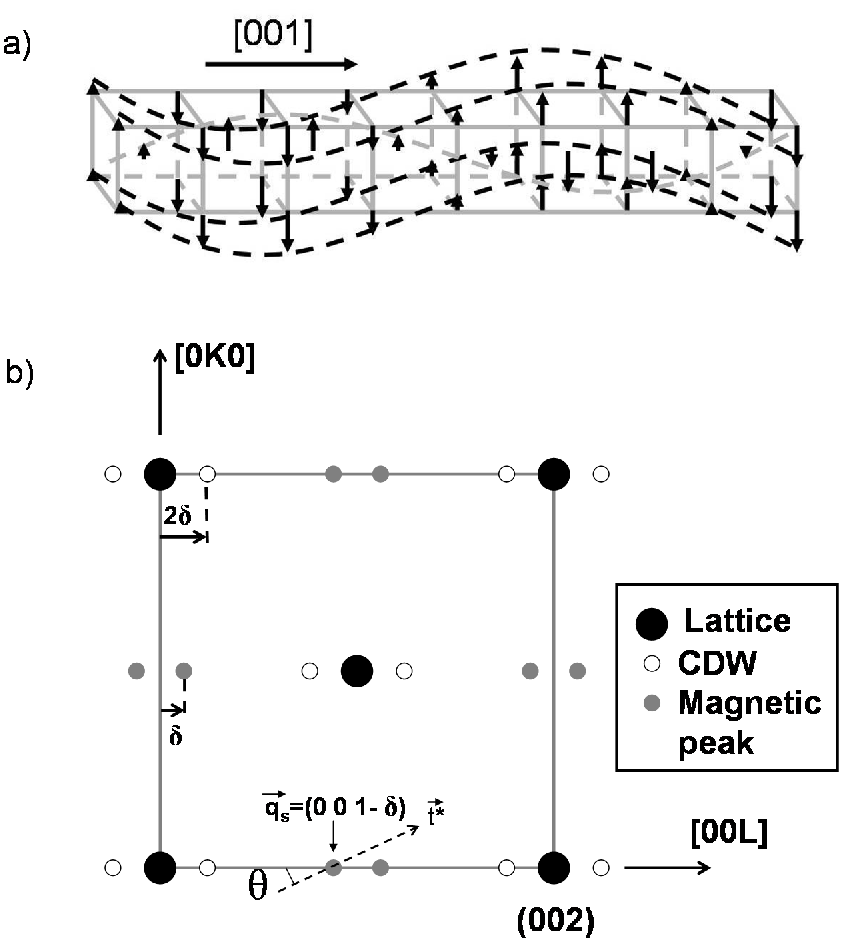}
}$$
\caption{a) Transverse SDW in chromium (real space) with SDW vector along [001]. 
For clarity, the SDW periodicity has been reduced to 7 cells instead of around 22. b) corresponding (100) scattering 
plane in the reciprocal space with
magnetic satellites reflections at $\vec{q}_s=(0\ 0\ 1-\delta)$ with $\delta=0.047$ r.l.u. at T=140K. 
At $\vec{q}_s$, the horizontal axis of the CCD camera cuts the (100) plane along $t^*$ (dashed line).}
\label{fig:1}       
\end{figure}

In diffraction experiments, magnetic satellites are observed at $\vec{q}_s=(0\ 0\ 1\pm\delta)$ in reciprocal space 
(see figure \ref{fig:1}b).
A CDW is also present in chromium giving rise to additional reflections at $2\vec{q}_s$ 
from bragg reflections~\cite{tsunoda,RefJ5}. More generally, several harmonics are visible. 
The odd harmonics are magnetic, whereas the even harmonics are charge components. 
The physical origin of this CDW is still controversal. At least two theories justify
 the appearance of these reflections~\cite{pynn}. The theory developped by Young and Sokoloff~\cite{youngsokoloff} 
is based on a 3 band calculation. In their approach, the existence of a second harmonic is due to the possible 
nesting between the two hole pockets (repulsive hole-hole interaction) leading to the CDW. This theory, based on 
nesting properties up to the second harmonic, explains the first order character of the magnetic transition~\cite{youngsokoloff}. \\
In another approach, the $2\vec{q_s}$ reflection was interpreted in~\cite{tsunoda} as due to 
the magnetostriction property of chromium, that couples strain to SDW. It was also shown that 
the spin flip transition at 123K could originate from strain effects~\cite{cowan}.

\subsection{Magnetic ordering in chromium}
\label{sec:13}

After years of intense research, the magnetic structure of chromium is now well understood \cite{RefJ3}.
The magnetic order in chromium~\cite{overhauser} can be described by the superposition of 
two waves on a host body-centered lattice \cite{Kachaturyan}: 
the antiferromagnetic (AF) order doubles the periodicity along the [111] direction, with a wave-vector $\vec{q_a}=(1 1 1)$, whereas the SDW develops along <100> directions, with a wave vector close to 
commensurability\footnote{We have chosen here the description of antiferromagnetism along $\vec{q_a}=(111)$ after~\cite{Kachaturyan} and~\cite{fishman2} but a description using the wave vector (100) is also possible.}. 
In general single-Q domains along any <100> direction are found in chromium but it is possible to grow single-Q single-domain samples, having a unidirectional SDW wave vector in the whole sample. The sample used was single-Q single-domain along the [001] direction because of the [001] orientation of the sample's surface \cite{cgiles}, with $\vec{q}_{s}=(0\ 0\ 1\pm\delta)$ and $\delta\approx0.047$ reciprocal lattice units (r.l.u.) at 140K (see figure~\ref{fig:1}.a). The absence of $(\pm\delta\ 0\ 1)$ and $(0\ \pm\delta\ 1)$ satellite reflections was checked during the experiment. From the position of the magnetic reflection in reciprocal space, the spatially dependent magnetic modulation can be written as:
\begin{equation}
\vec{\mu}(\vec{r}) = \vec{\mu_M}\cos(\vec{q_s}\cdot\vec{r}+\phi(\vec{r})),
\label{eqn1}
\end{equation}
where $\phi$ is a phase, dependent on the position $\vec{r}$. After expansion of this cosine function, and taking into account the atomic structure of chromium, the two contributions of the magnetic modulation can be separated, and $\vec{\mu}(\vec{r})$ takes the form: 
\begin{equation}
\vec{\mu}(\vec{r}) \propto \cos(\vec{q_a}\cdot\vec{r}+\phi_a(\vec{r}))\times\cos(2\pi\delta\cdot z+\phi_s(\vec{r})),
\label{eqn2}
\end{equation}
where $\phi_a(\vec{r})$ and $\phi_s(\vec{r})$ are the antiferromagnetic and SDW phases respectively. The first cosine function describes the commensurate antiferromagnetism found along the [111] direction and the second one the SDW modulation along the [001] direction. This notation involving two uncoupled modulations is convenient to separate the spatially dependent phases $\phi_a$ and $\phi_s$ considered later in this paper. Their spatial dependence will be used to describe a magnetic dislocation.

\section{Experiment}
\label{sec:2}
By tuning the incident x-rays to below the K-edge of Cr (5.989 keV)~\cite{RefJ6}, a favorable condition for coherent x-ray diffraction experiments can be found. At this energy, it is possible to observe the magnetic satellites by non resonant magnetic scattering and to achieve a relatively large transverse coherence length. 

In this section, we will describe both non-resonant magnetic scattering and coherent x-ray scattering before introducing the experimental setup.

\subsection{Coherent x-ray diffraction: longitudinal and transverse coherence length}
\label{sec:2.1}
In coherent diffraction experiments, one has to take into account both 
the longitudinal and the transverse coherence lengths ($\xi_L$ and $\xi_t$ respectively) of the x-ray beam.
The longitudinal coherence length $\xi_L$ is governed by the bandpass of the monochromator 
and by the wavelength $\lambda$ of the incident beam as:
$$\xi_L=\frac{\lambda^2}{2\Delta\lambda}$$
In our case, $E=5.9$ keV ($\lambda=2.1$ $\mathring{A}$) and $\frac{\Delta\lambda}{\lambda}\approx10^{-4}$. 
This leads to a longitudinal coherence length $\xi_L=1.05$ $\mu m$. Taking into account the large penetration depth of the x-ray beam 100 eV below the K-edge of chromium ($\mu^{-1}=20$ $\mu m$), the path length difference calculated for the SDW scattering angle ($\theta_{SDW}=20.36^o$) gives: $2\mu^{-1}\sin^2{\theta}=4.8$ $\mu m$. It is more than 
four times as big as $\xi_L$ so a weak visibility is expected along the longitudinal direction.

Optical aberrations, and more precisely the quality of the surface of the x-ray mirrors, are the main reason for a decrease of the transverse coherence length. More precisely, the low and high frequencies of a reflecting surface (slope errors and roughness respectively) alter the transverse coherence quality in quite a different ways.
If the roughness of a mirror is described by a gaussian distribution with a standard deviation 
$\sigma$, its contribution will attenuate the whole diffraction pattern by a factor $\exp{[-(4\pi\sin\alpha)^2(\sigma/\lambda)^2]}$ where $\alpha$ is the grazing incident angle on the mirror. 
Similar to a Debye Waller factor, this contribution only attenuates the global profile of the diffraction pattern. Unlike visible light, x-rays are thus more sensitive to surface roughness but this attenuation remains limited because of the small $\alpha$ value and the fact that the wavelength ($\lambda=2.1$ $\mathring{A}$) is of the same order of magnitude than typical surface roughness of modern mirrors. 

On the other hand, slope errors of the mirror may induce dramatic effects on the coherent diffraction patterns.
 Interferences may occur between those correlations and those of the sample which are difficult to evaluate.

In the worse case, 
if one considers that the transverse degree of coherence is zero at the exit of the optical part of the beamline, because of optical aberrations, the transverse coherence length $\xi_t$ can be written as:
\begin{equation}
\xi_t=\frac{\lambda}{2}\frac{R}{S},
\label{xit}
\end{equation}
where $R$ is the optics-sample distance and $S$ the secondary source aperture.
In our experiment, this leads to $\xi_t\approx5$ $\mu m$ in the vertical direction and 2~$\mu m$ in the horizontal direction.
\begin{figure}[!ht]
$$\resizebox{0.75\columnwidth}{!}{%
\includegraphics[width=\columnwidth]{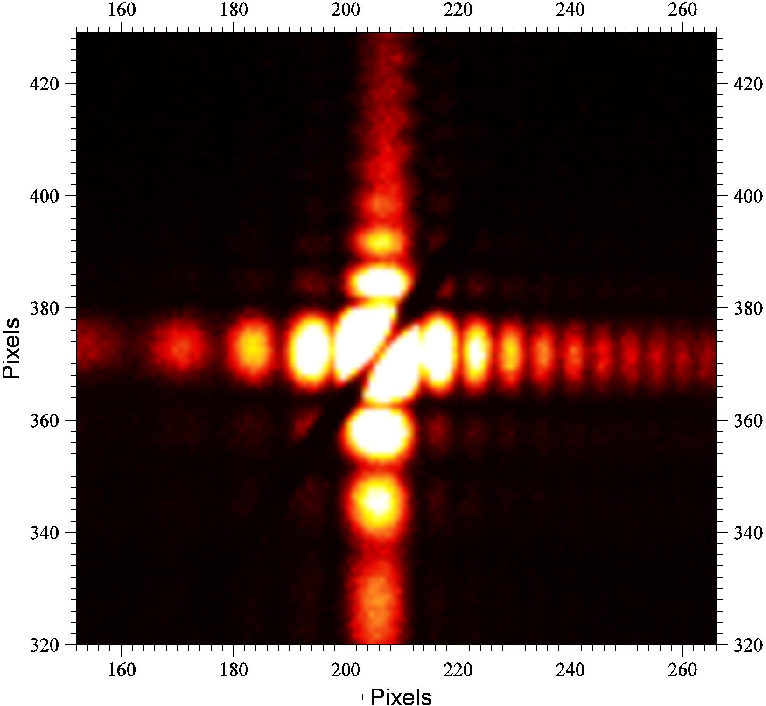}
}$$
\caption{(Color online) Interference fringes obtained at 8keV (Log scale). 
Pinholes are closed at $2\mu m\times2\mu m$ and the $22\mu m$-pixel-size CCD
 camera is located 2,22m downstream. A diagonaly placed absorbing wire 
has been used across the direct beam to protect the camera. Cross terms are also visible. }
\label{fig:2}       
\end{figure}
 As a preliminary test of the transverse coherence length, we have measured the diffraction pattern of $2\mu m \times 2\mu m$ square slits using an $8keV$ beam (figure~\ref{fig:2}).
 This 'pinhole', located $22cm$ upstream the sample position,
 was made by two perpendicular sets of tungsten blades. The x-rays were detected by a CCD camera cooled to $-50^oC$, positioned 2.2m downstream of the slits. The pixel size of the CCD was $22\mu m\times22\mu m$ which gave a resolution in reciprocal space of $\Delta q=0.56\cdot10^{-4}$ $\mathring{A}^{-1}$ in the radial direction. The regularity and high contrast of the fringes\footnote{Note here that an asymmetry of the diffraction pattern is observed because the blades of the diffracting slits are not coplanar \cite{coherence}.} give evidence of the large transverse coherence length.\\

\subsection{Non-resonant x-ray magnetic diffraction}
\label{sec:2.2}
Non-resonant magnetic scattering is highly sensitive to the incident x-ray polarization and can result in full or partial rotation of the scatttered x-rays' polarization. We can consider the incident x-ray to be either $\sigma$ or $\pi$ polarized, corresponding to the case where the electric field vector of the x-ray is perpendicular or parallel to the scattering plane respectively. Likewise, the scattered x-ray can be $\sigma'$ or $\pi'$ polarized. The resulting polarization is dependent on both the orbital angular momentum $\mathscr{L}$ and the spin components $S_x$, $S_y$ and $S_z$~\cite{blume}. Non-resonant scattering has been previously used to study Cr at an incident energy 100 eV below the Cr K-edge~\cite{RefJ6,RefJ4}.

In our case, the incident beam is $\pi$ polarized in the horizontal diffraction plane (see figure~\ref{fig:3}). 
Since no polarization analysis of the scattered beam is performed, the magnetic scattering amplitude is the sum of 
the two  $\pi\sigma'$ and $\pi\pi'$ scattering contributions, $M_{\pi\sigma'}$ and $M_{\pi\pi'}$ respectively: 
\begin{eqnarray*}
A(\theta)&=&-i\frac{\hbar\omega}{mc^2}r_0\left(M_{\pi\pi'}+M_{\pi\sigma'}\right)\\
&=&-i\frac{\hbar\omega}{mc^2}r_0\left(S_y\sin{2\theta}-2S_x\sin^2{\theta}\cos{\theta}\right)
\end{eqnarray*}
where $\hbar\omega$ is the incident photon energy, $r_0$ the classical electron radius, 
$m$ the mass of the electron and $c$ the light velocity. There is no contribution of $\mathscr{L}$ and $S_z$ in this expression because $\mathscr{L}=0$ in chromium (experimentally shown in~\cite{RefJ4}), and we worked in the transverse polarization regime for which $S_z=0$.

\subsection{Experimental setup}
\label{sec:2.3}

\begin{figure}[!ht]
$$\resizebox{0.75\columnwidth}{!}{%
\includegraphics[width=\columnwidth]{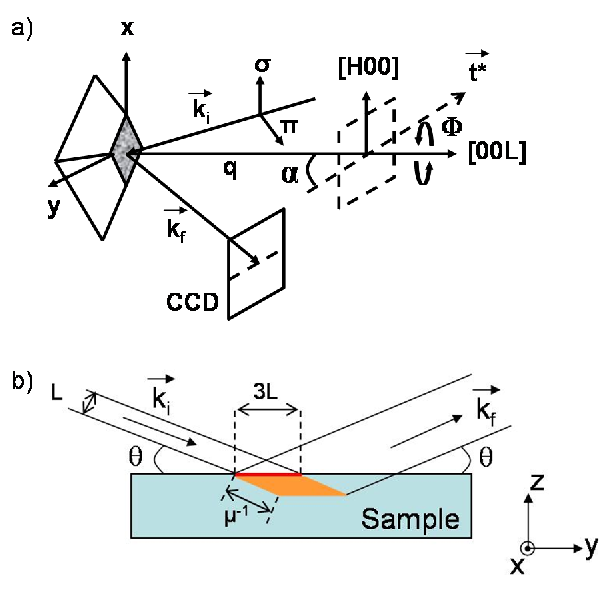}

}$$
\caption{(Color online) a) Experimental setup. $\Phi$ is the azimuth angle, $\alpha$ is the angle between 
the CCD camera and the $[00L]$ direction in reciprocal space. $\sigma$ and $\pi$ polarizations are displayed; b) Projection of the beam on the surface of the sample. $\mu^{-1}=20\mu m$ is the penetration depth, $\theta$ the incident angle of the beam, and $L$ is the beam size.}
\label{fig:3}       
\end{figure}

The high-quality single crystal had a truncated pyramidal shape with a square base of $36mm^2$, a top square platform of $2.25mm^2$ 
and was about $1.5mm$ high (see figure~\ref{fig:3}). 
Though the SDW is already present at room temperature, the sample 
was cooled to 140K to reach the maximum of the intensity of the magnetic reflection \cite{RefJ3}. 
A top-loading He flow cryostat was used, with a temperature stability of $1mK$. The experiments were carried out at the 
European Synchrotron Radiation Facility (ESRF), at the magnetic scattering beamline ID20~\cite{luigi}.
 The first harmonic of two undulators (U32 and U35) was used and a Si(111) monochromator provided 
a monochromatic beam of x-rays at 5.9keV. This energy was chosen to eliminate fluorescence from the sample. 
At this energy, only a few reflections were 
available because of the small lattice parameter ($a=2.88\mathring{A}$). The direct beam was found to be limited with a Full Width at Half Maximum (FWHM) no larger than 1 pixel of the 22$\mu m$-pixel-size CCD camera. Indeed, the beam was weekly divergent because we were not fully in the Fraunhofer regime~\cite{discuss_fraunhofer}. The $\vec{q}_s=(0\ 0\ 1-\delta)$ magnetic satellite 
($\theta_s$ = 20.36$^o$) was measured along with the (002) fundamental reflection ($\theta_{002}$=46.8$^o$) 
in reflection geometry.
 
Although the (002) wave vector was almost parallel to the polarization of the incident beam ($2\theta_{002}$ close to $90^o$), the (002) reflection was intense ($\approx$$2.5\cdot10^9$ cts/s). 

The major difficulty in this experiment was dealing with the weak intensity of the $\vec{q}_{s}$ magnetic satellite. 
The non-resonant magnetic scattering cross-section is intrinsically weak, and this was exacerbated by the need for a high degree of coherence which imposed pinhole sizes of a few microns, leading to a major loss of intensity. Moreover, the value of the azimuth $\Phi$ (see figure~\ref{fig:3}) was not controlled inside 
the cryostat, and thus not optimized to have the maximum of the intensity \cite{RefJ4}. 
However, all this was partially compensated by the fact that the sample displayed a single SDW domain, that the beamline provided a highly brilliant beam (two undulators are used in series \cite{luigi}), and that the pinholes were opened
 at twice the value generally used in coherence experiments (generally $10\mu m$ at $\lambda$=1.54$\mathring{A}$). 
This last point allowed us to gain a factor 4 on the intensity keeping a good total transverse coherence length
thanks to the large wavelength used here (see eq.\ref{xit}). With this experiment setup, we obtain an acceptable degree of coherence ($\beta=18.5\%$ as defined in \cite{livet}). We finally obtained approximately 0.7 cts/s on the $\vec{q}_{s}$ satellite reflection.
All diffraction patterns shown in this paper are the sum of 180 10s-acquisition and are obtained after treatment \cite{livet2}.

At the magnetic reflection, the vertical axis of the CCD camera corresponds to the [H00] axis 
in reciprocal space, and the horizontal one makes a $20^o$ angle with the [00L] radial direction (see figures~\ref{fig:1}.b. and~\ref{fig:3}.a.). 
This last direction will be called $\vec{t}^*$ in the following.
A slight contribution of the [0K0] axis is thus present on the horizontal direction of the CCD.

\section{Results - SDW profiles}
\label{sec:2.5}
\begin{figure}[!ht]
$$\resizebox{\columnwidth}{!}{%
\includegraphics[width=\columnwidth]{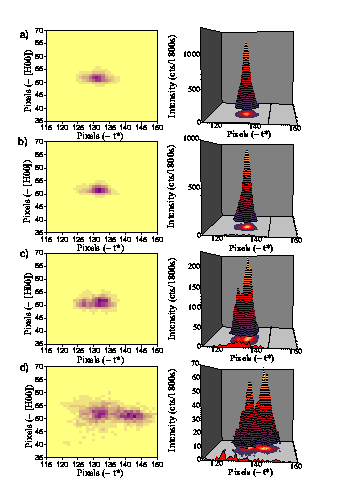}
}$$
\caption{(Color online) Magnetic reflection observed at $T=130K$ 
at four beam positions aligned along the $y$-axis with $20\mu m$ steps. The horizontal axis
corresponds to the $\vec{t}^*$ direction defined in the text. 
The vertical axis corresponds to the [H00] direction.
The 3D plots on the right hand side are continuous extrapolations of corresponding pixelated images on the left hand side.}
\label{sdw}       
\end{figure}
 For most beam positions, the magnetic reflection displays a single peak as in figure~\ref{sdw}.a. The peak is elongated along $\vec{t}^*$, with a FWHM spreading over 6 pixels in that direction, corresponding to a real space size of $3\mu m$, which has been estimated from simulations. In fact, the width of the satellite along this radial direction is driven by the penetration length perpendicular to the sample surface: $\frac{1}{2}\mu^{-1}\sin{\theta}\approx3.5\mu m$ \cite{robinson} (see figure~\ref{fig:3}.b.). Thus, the observed broadening along $\vec{t}^*$ is due only to the finite x-ray penetration. We therefore conclude that the diffraction pattern corresponds to a single magnetic domain.\\
When the beam is moved $20\mu m$ away on the sample's surface along the $y$-axis, 
the reflection continuously splits and takes an asymetric shape (figure~\ref{sdw}.c). When the beam is moved again by $20\mu m$, the magnetic reflection is split into two parts having almost the same intensity and same width. Each part of the split reflection has the same width as the single magnetic reflection in figure~\ref{sdw}.a (see also the corresponding profiles in figure~\ref{fits}).

\section{Interpretation}
\label{sec:3}

Those pictures are very different from speckle patterns due to any disorder observed by coherent 
x-ray diffraction at wide angles (see for example \cite{srtio}). The two speckles observed along 
the horizontal axis in figure~\ref{sdw}.d can not be due to two magnetic domains having different
 orientations because this axis of the camera cuts the reciprocal lattice mainly along the radial direction. 
The presence of several magnetic domains having different orientations would give rise to multiple peaks in 
the transverse direction. \\
It cannot be explained either by the presence of two domains in the radial direction, with slightly different wave vectors $\vec{q}_s$, 
because the two spots on figure~\ref{sdw}.d would be twice as broad as the single peak of figure~\ref{sdw}.a. To our point of view, 
this diffraction pattern is consistent with a phase shift of the magnetic modulation. 

As in CDW systems \cite{RefJ1}, the SDW is defined by an unidimensional wave vector. In chromium however, the situation is slightly more complicated because
the SDW is coupled with a non-parallel AF modulation. There are thus two ways to describe the same phase shift of the magnetic order of chromium, 
either with respect to the SDW or the AF order, considering $\phi_s(\vec{r})$ or $\phi_a(\vec{r})$ respectively (see equation~\ref{eqn2}). For example, a pure screw dislocation of the SDW corresponds to a mixed dislocation (between a screw and an edge)
of the AF modulation. In the following, we will consider only spatial variations of $\phi_s(\vec{r})$ to describe any 
dislocation of the magnetic order in chromium.
 As an example, an edge dislocation whose line is along the x-axis is built first by introducing a $\Delta\phi_s=\pi$ phase shift between the $y>y_0$ and $y<y_0$ half-planes, and then connecting  the wave fronts with the continuous function $arctg\left(\frac{z-z_0}{y-y_0}\sqrt{\frac{K_y}{K_z}}\right)$ \cite{friedel2}. $K_y$ and $K_z$ are the SDW force constants along $y$ and $z$ respectively, and ($y_0$,$z_0$) determines the position of the dislocation line in the volume. Three parameters are necessary to determine the topology of the defect: the SDW force constants ratio $\sqrt{\frac{K_y}{K_z}}$, and the cartesian coordinates of the position of the dislocation line ($y_0$,$z_0$). 
 
\section{Simulations}
 
Simulations have been carried out in order to reproduce the observed diffraction patterns, taking into account the [111] AF and the [001] SDW modulation of chromium. The simulation box was taken cubic ($L\times L\times L$) with $L=60$ cells. L has been chosen large enough in such a way that truncation effects do not
influence the shape of diffraction patterns. The widths and intensities of the simulated reflections have been rescaled to fit the experimental data. Note that a change in the volume dimension $L$ only modifies the widths of the reflections but not their profile, which justifies this rescaling.

\begin{figure} 
 $$\resizebox{0.95\columnwidth}{!}{%
\includegraphics[width=\columnwidth]{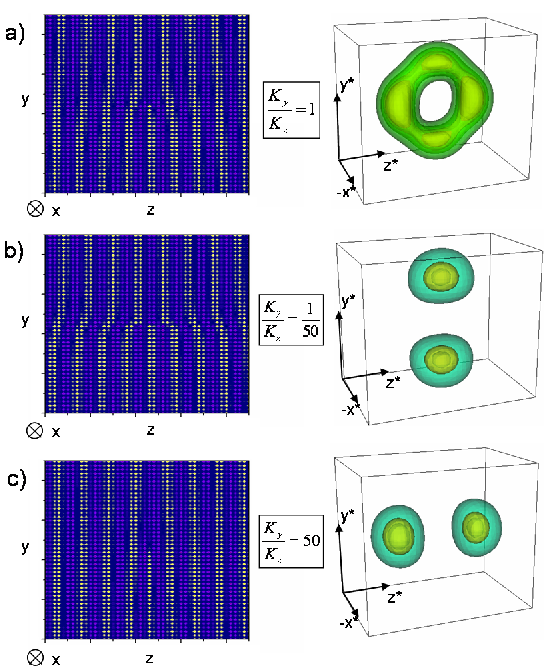}
}$$
\caption{(Color online) Effect of the SDW force constants on the shape of the reflection in reciprocal space. On the left hand side, a dislocation of the SDW in chromium is represented in real space (the represented SDW has a periodicity 11 instead of 22), and on the right hand side, the associated magnetic reflection is shown. Each atom in real space is represented by a dot and the color range is related to the amplitude of the magnetic moments. a) Isotropic case $\sqrt{\frac{K_y}{K_z}}=1$: the reflection in reciprocal space is characterized by four maxima in the plane perpendicular to the direction of the dislocation line. b) $\sqrt{\frac{K_y}{K_z}}=1/50$ ; c) $\sqrt{\frac{K_y}{K_z}}=50$. We can clearly see from b) and c) that anisotropic force constants favor one unique direction for the splitting.}
\label{dislogeneral}
\end{figure}

In general, the force constants of density waves can be obtained from dispersion curves measured by inelastic x-ray~\cite{phason} or neutron scattering. In chromium, the longitudinal dispersion curve
has been measured in \cite{dispersion} but as far as we know, no measurement has been carried out along the transverse direction. The relaxation around any phase shift is thus unknown {\it a priori}. One of purpose of this work is to show that the ratio $\sqrt{\frac{K_y}{K_z}}$ can be estimated from the shape of the diffraction pattern. Indeed, in an isotropic case ($K_y=K_z$), the presence of a dislocation and the relaxation around it has a huge impact on the shape of a reflection, as shown on figure~\ref{dislogeneral}.a: it splits into four parts in the plane perpendicular to the dislocation line. When the force constants are anisotropic, strong variations are induced on the distribution of intensity around the magnetic reflection. Such a change in the force constants ratio favours one direction for the splitting, keeping the centro-symmetry of the image (figures~\ref{dislogeneral}.b and c).

The other relevant parameter is the position of the dislocation line in the probed volume. If the dislocation line is in the center of the volume, the two spots of a split reflection have the same intensity. If it is moved from the center, the centro-symmetry of the intensity distribution is broken: the two spots no longer have equal intensities.

\begin{figure} 
$$\resizebox{0.95\columnwidth}{!}{%
\includegraphics[width=\columnwidth]{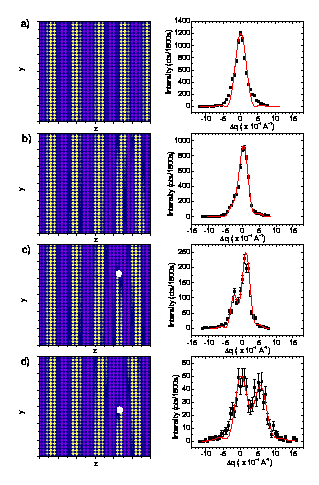}
}$$
\caption{(Color online) On the right hand side are displayed the experimental profiles (black squares) through the magnetic reflection along $\vec{t}^*$, for several beam positions along the $y$-axis. Those profiles
 correspond to a slide of the 2D patterns in figure~\ref{sdw}. Simulated profiles along the same direction are displayed
with continuous lines. They correspond to the real space configuration displayed in the (100) plane on the left hand side. a) Magnetic order in chromium without defect which leads to a single magnetic reflection; b)c)d) Same magnetic order containing an edge dislocation 
(the dislocation line is shown by a white dot), with $\sqrt{\frac{K_y}{K_z}}=7$. The splitting of the magnetic reflection becomes more and more symmetric as the dislocation line is moved towards the center of the probed volume.}
\label{fits}
\end{figure}

 During the experiment, the beam was moved by steps of 20$\mu m$, corresponding to the beam size.
 The dependence of the diffraction patterns versus the beam positon along $y$ has been a strong constraint which
significantly reduced the number of possible solutions.
In this work, we considered the cases of an edge, a screw and a mixed dislocation. 
Among every tested solutions, an edge dislocation whose 
line develops along the $x$-axis, with $\sqrt{\frac{K_y}{K_z}}=7$ reproduces the most correctly 
the observed splitting in figure~\ref{sdw}.d. Such a dislocation induces a splitting  along the [0K0] and [00L] directions. 
A split satellite is observed (see figure~\ref{fits}.d) because the CCD camera cuts the reciprocal 
lattice along $\vec{t}^*$, which is a combination of these two directions. The symmetric splitting
 is well reproduced when the dislocation line is located very close to the middle of the probed volume 
in agreement with figure~\ref{sdw}d.
 
 The continuous disappearance of the splitting as the beam moves along $y$ (\ref{sdw}.c and \ref{sdw}.b) 
 can be well simulated by shifting the dislocation line from the center of the volume (\ref{fits}.c and \ref{fits}.b):
  the three 
  images shown in figure~\ref{sdw}.b, \ref{sdw}.c and \ref{sdw}.d can be well fitted with a dislocation line positionned at different values along 
  the $y$-axis (see figures~\ref{fits}b, \ref{fits}c and \ref{fits}d). 
  The splitting disappears when the dislocation line is too far from the center (figure~\ref{fits}a).

  The best fit was obtained by successive translations of $18\mu m$ (taking into account the scale factor corresponding to the simulation box size L=60), which is in good agreement with the experimental value of $20\mu m$.
  
A SDW \textit{screw} dislocation developing along the same axis with the same force constant ratio can also 
be fitted well with our measurements. It is interesting to note here that the direction of the dislocation line affects
 the shape of the pattern more than the nature of the defect does. Our measurements do not allow to discriminate between an edge or a screw dislocation.

\section{Discussion}
\label{sec:4}

Several diffraction patterns have been obtained for different beam positions along the $y$-axis.
They
are consistent with the presence of a single magnetic dislocation embedded at few micrometers from the surface. 
Such a phase shift of the SDW can reproduce the main features of the observed data:
1) the same width for the two peaks of the split magnetic reflection, which is also the same width as the single reflection, and 
2) the continuous splitting of the magnetic reflection as respect to the beam position, mainly along the radial direction.
However, the intensity of the split reflection on figure~\ref{fits}.d is smaller than what we expect. Besides, the reflection on this image is not exactly at the same angle as the others, which suggests that the SDW wave fronts are slightly distorted in the close vicinity of the dislocation line. We insist on the fact that mosaicity of the magnetic order or the presence of domains with different wave vectors can not explain these features.\\
It has been shown that coherent x-ray diffraction allows to measure strain fields induced by the presence of individual dislocations~\cite{irobinson}. We show in this paper that another powerful feature of these measurements is that they allow to extract a value for the SDW force constants ratio. Comparison with CDW systems is interesting. In quasi-1D systems, the CDW is stiffer parallel to the CDW wave vector. Therefore, in such compounds as blue bronze, the diffraction pattern in presence of a dislocation presents a splitting perpendicularly to the chains axis $b^*$ \cite{RefJ1}. Our measurements show that in chromium the splitting appears along the SDW wave vector. This is why the force constants ratio found in our simulations for chromium $\sqrt{\frac{K_y}{K_z}}=7$ is inverse than in blue bronze ($\sim 1/10$). Inelastic neutron scattering experiments confirmed the value for blue bronze CDW in~\cite{hennionpouget}. Such measurements have not been carried out on the SDW of chromium. However, many differences between chromium and blue bronze exist and could explain the difference concerning the force constants ratio. The first difference concerns the atomic structure of the two compounds: chromium has a 3D isotropic structure and its SDW, originating from delocalized electrons, is very different from the strongly lattice-coupled CDW found in quasi-1D systems. Moreover, magnetic interactions are involved in chromium and not in CDW systems. Besides, chromium develops additional magnetic excitations, and these so-called wavons also induce a slight anisotropy in the SDW dispersion curve as predicted theoretically in \cite{theory}. This theoretical prediction gives a force constant ratio that is less significant ($\sqrt{\frac{K_y}{K_z}}\sim2$) than the estimations given by the presence of a dislocation in our work. But in both cases, the SDW is found to be stiffer in the transverse direction than in the radial one. In all the cases, other experiments are needed to confirm and explain the magnetic force constants in chromium.\\
A SDW screw dislocation perpendicular to the (001) surface observed by spin-polarized STM technique has already been reported in~\cite{kleiber}. It is important to note that the dislocation found in our work is very different from that screw dislocation. In their case, the screw dislocation was induced by a step at the surface.\\

\section{Conclusion}
To conclude, we show in this paper that it is possible to probe a magnetic defect embeded in the bulk using coherent non-resonant magnetic scattering. Our data are consistent with a magnetic dislocation, and the estimated SDW force constants are found to be anisotropic. The value found for the force constants ratio goes in the same way as the one predicted theoretically in \cite{theory}. Beyond the case of chromium, this work shows that coherent diffraction combined with magnetic scatering may be
 applied to the study of bulk magnetic phase defects.\\
The authors would like to aknowledge A. Thiaville, J.P. Pouget, J.P. Jamet, C. Pasquier and N. Kirova for fruitful discussions.

\end{document}